# Computational fluid dynamics modelling of left valvular heart diseases during atrial fibrillation

Stefania Scarsoglio[1], Andrea Saglietto[2], Fiorenzo Gaita[2], Luca Ridolfi[3] and Matteo Anselmino[2]

[1] Department of Mechanical and Aerospace Engineering, Politecnico di Torino, Torino, Italy
[2] Division of Cardiology, Department of Medical Sciences, "Città della Salute e della Scienza" Hospital, University of Turin, Torino, Italy
[3] Department of Environmental, Land and Infrastructure Engineering, Politecnico di Torino, Torino, Italy

## ABSTRACT

**Background:** Although atrial fibrillation (AF), a common arrhythmia, frequently presents in patients with underlying valvular disease, its hemodynamic contributions are not fully understood. The present work aimed to computationally study how physical conditions imposed by pathologic valvular anatomy act on AF hemodynamics.

**Methods:** We simulated AF with different severity grades of left-sided valvular diseases and compared the cardiovascular effects that they exert during AF, compared to lone AF. The fluid dynamics model used here has been recently validated for lone AF and relies on a lumped parameterization of the four heart chambers, together with the systemic and pulmonary circulation. The AF modelling involves: (i) irregular, uncorrelated and faster heart rate; (ii) atrial contractility dysfunction. Three different grades of severity (mild, moderate, severe) were analyzed for each of the four valvulopathies (AS, aortic stenosis, MS, mitral stenosis, AR, aortic regurgitation, MR, mitral regurgitation), by varying–through the valve opening angle–the valve area.

**Results:** Regurgitation was hemodynamically more relevant than stenosis, as the latter led to inefficient cardiac flow, while the former introduced more drastic fluid dynamics variation. Moreover, mitral valvulopathies were more significant than aortic ones. In case of aortic valve diseases, proper mitral functioning damps out changes at atrial and pulmonary levels. In the case of mitral valvulopathy, the mitral valve lost its regulating capability, thus hemodynamic variations almost equally affected regions upstream and downstream of the valve. In particular, the present study revealed that both mitral and aortic regurgitation strongly affect hemodynamics, followed by mitral stenosis, while aortic stenosis has the least impact among the analyzed valvular diseases.

**Discussion:** The proposed approach can provide new mechanistic insights as to which valvular pathologies merit more aggressive treatment of AF. Present findings, if clinically confirmed, hold the potential to impact AF management (e.g., adoption of a rhythm control strategy) in specific valvular diseases.

## INTRODUCTION

Atrial fibrillation (AF) is the most prevalent sustained tachyarrhythmia, currently affecting up to 2% of the general population (*Andrade et al., 2014*), producing symptoms (such as chest pain, palpitations, reduced exercise tolerance, shortness of breath) and decreasing cardiac performance (*Fuster et al., 2006*). With an estimated number of 33.5 million individuals affected worldwide in 2010, AF has almost reached epidemic status (*Piccini & Daubert, 2014*) and is becoming a public health problem in developing countries (*Nguyen, Hilmer & Cumming, 2013*). Therapeutic approaches can either pursue rhythm control–i.e., restoring and maintaining sinus rhythm by antiarrhythmic drugs or transcatheter ablation–or rate control along–i.e., reducing ventricular rate to reduce symptoms and improve quality of life (*January et al., 2014*).

Even though previous clinical data, such as those resulting from the AFFIRM trial (*Wyse et al., 2002*), suggested that rate control is not inferior to rhythm control in terms of survival advantages, this topic is still widely debated and questioned (*Al-Khatib et al., 2014*; *Ionescu-Ittu et al., 2012*). In fact, current literature primarily refers to AF patients in general, without focusing on the concomitant effect of underlying valvular disease present in a relevant subgroup of AF patients (*Darby & DiMarco, 2012*; *Vora, 2006*). In addition, hemodynamic measurement data are limited, as AF patients with valvular diseases are usually excluded from clinical trials so most data are restricted to echocardiographic measurements (*Dahl et al., 2014*; *Kristensen et al., 2012*). Moreover, interest often focuses on postoperative effects of valve surgery for AF patients (*Fukunaga et al., 2008*; *Lim et al., 2001*).

AF and valvular diseases are often present simultaneously, however their relative hemodynamic contributions remain unclear (*Levy, 2002*; *Molteni et al., 2014*). Although AF is widely recognized as a risk marker for valve diseases (*Gertz et al., 2011*; *Enriquez-Sarano & Sundt, 2010*; *Levy et al., 2015*) and is responsible for aggravating valvulopathies already present (*Grigioni et al., 2002*; *Dujardin et al., 1999*; *Yamasaki et al., 2006*), in clinical practice it is not easy to understand how physical limitations induced by valvulopathies act on hemodynamics in AF. In fact, discerning which changes are due to altered valvular dynamics and which are related to the arrhythmia is rather difficult, and therefore the overall hemodynamic response in the presence of both pathologies is usually studied. Moreover, some measurements, such as those based on peak inflow velocity, are not reliable to study the role of the valvulopathy during AF (*Özdemir et al., 2001*; *Thomas, Foster & Schiller, 1998*). From a computational perspective, mathematical modelling offers new insights into the dynamics of valvular diseases and their effects on the whole cardiovascular system (*Mynard et al., 2012*; *Broomé et al., 2013*; *Domenichini & Pedrizzetti, 2015*). However, to the best of our knowledge, the concomitant presence of AF and left heart valvulopathies has not been analyzed to date.

A computational approach in this scenario aims to overcome the aforementioned gaps. The effects of valve pathology and its severity in presence of AF were studied and compared, from a fluid dynamics point of view, with respect to a reference configuration where AF is present in the absence of valvular pathology (lone AF). Based on a lumped-parameter model of the cardiovascular system validated during AF conditions and characterized by a customizable valve dynamics (*Scarsoglio et al., 2014*; *Anselmino et al., 2015*; *Scarsoglio et al., 2016*), we simulated hemodynamics in AF with different grades of left-sided valvular diseases (aortic stenosis, AS; mitral stenosis, MS; aortic regurgitation, AR; mitral regurgitation, MR) to elucidate the hemodynamic consequences that they produce during AF. Simulations were carried out over thousands of heart beats, therefore ensuring the statistical stationarity of the results. Simultaneous hemodynamic parameters can be derived without approximating, since the complete temporal series of the cardiovascular variables (pressure, volume, flow rate) were obtained as the primary output of the model. Moreover, specific severities of valvular pathology can be evaluated, by mathematically relating the valve opening angle and the valve area, according to the current guidelines for valve diseases (*Baumgartner et al., 2009*; *Lancellotti et al., 2010a*; *Lancellotti et al., 2010b*).

This study, concerning a somewhat surprisingly neglected topic, provides new insights into valvular heart diseases during AF, potentially suggesting which valvular diseases, from a computational hemodynamic point of view, might require more aggressive AF management (e.g., a rhythm control strategy such as AF transcatheter ablation). Our modelling outcomes revealed that both mitral and aortic regurgitation strongly affect hemodynamics, immediately followed by mitral stenosis, while aortic stenosis has the least impact among the analyzed valvular diseases.

## MATERIALS AND METHODS

### Cardiovascular model, variables and parameters definition

The cardiovascular model used here, first proposed by *Korakianitis & Shi (2006)* for healthy and diseased valves, has then been validated over more than 30 clinical measurements regarding AF (*Scarsoglio et al., 2014*). It has been recently adopted to evaluate, from a computational point of view, the impact of higher HR during AF at rest (*Anselmino et al., 2015*), as well as the role of AF in the fluid dynamics of healthy heart valves (*Scarsoglio et al., 2016*).

The model relies on a lumped parameterization of the four heart chambers, together with the systemic and pulmonary circulation. Cardiac and circulatory regions are described using electrical terminology, such as compliance (accounting for the elastic properties), resistance (simulating the viscous effects) and inductance (approximating inertial terms). The resulting ordinary differential system is expressed in terms of pressure, $P$ [mmHg], volume, $V$ [ml], flow rate, $Q$ [ml/s], and valve opening angle, $\vartheta$ [°]. Each of the four heart chambers is active and governed by an equation for mass conservation (considering the volume variation), a constitutive equation (for the pressure-volume relation through a time-varying elastance, $E$), an orifice model equation (relating pressure and flow rate), and an equation for the valve motion mechanisms. Both systemic and pulmonary circuits are partitioned into four arterial and one venous sections. Each circulatory compartment is ruled by an equation for mass conservation (in terms of pressure variation), an equation of motion (flow rate variation) and a constitutive linear equation between pressure and volume. The elastic vessel properties are in general dependent on the pressure level. However, a linear relation between pressures and volumes can be assumed in the range of physiological values (*Ottesen, Olufsen & Larsen, 2004*). The complete system was numerically solved through an adaptive multistep scheme implemented in Matlab. Since the cardiovascular dynamics present stiff features, i.e. rapid and abrupt variations in time, a stiff solver implemented in the *ode15s* Matlab function was adopted (all the modeling and computational details are given in *Scarsoglio et al. (2014)*).

We focused here on the left heart dynamics by means of pressure ($P$) and volume ($V$) variables, also evaluating end-diastolic (ed) and end-systolic (es) values: left atrial pressure and volume ($P_{la}$ and $V_{la}$, respectively), left ventricle pressure ($P_{lv}$) and volume ($V_{lv}$, $V_{lved}$, $V_{lves}$), systemic arterial pressure ($P_{sas}$, $P_{sas,syst}$, $P_{sas,dias}$), pulmonary arterial ($P_{pas}$) and venous ($P_{pvn}$) pressures. End-systole is the instant defined by the closure of the aortic valve, while end-diastole corresponds to the closure of the mitral valve. We introduce $RR$ [s] as the temporal range between two consecutive heart beats, while $HR$ [bpm] is the heart rate, i.e., the number of heart beats per minute. Performance indexes are computed as well:

- stroke volume, $SV = V_{lved} - V_{lves}$ [ml];
- ejection fraction, $EF = SV/V_{lved} \times 100$ [%];
- cardiac output, $CO = (FV_{ao} + RV_{ao}) \times HR$ [l/min], where $FV$ [ml/beat] and $RV$ [ml/beat] are the forward and regurgitant volumes, respectively. The forward volume

$$FV = \int_{RR} Q^+(t) dt, \tag{1}$$

is the volume of blood per beat flowing forward through the valve (the symbol $Q^+$ indicates the positive flow rate outgoing from the valve), while the regurgitant volume

$$RV = \int_{RR} Q^-(t) dt, \tag{2}$$

is the volume of blood per beat which regurgitates backward through the valve, with the symbol $Q^-$ representing the negative flow rate going backward through the valve ($RV < 0$ by definition). As $FV$ and $RV$ are here computed for the aortic valve, $FV_{ao} + RV_{ao}$ is the net volume per beat [ml/beat] across the aortic valve (*Scarsoglio et al., 2016*).

## Valve dynamics

The valve dynamics introduced by *Korakianitis & Shi (2006)* include several mechanisms, such as the pressure difference across the valve, the dynamic motion effect of the blood acting on the valve leaflet, the frictional effects from neighboring tissue resistance and the action of the vortex downstream of the valve. Only the shear stress on the leaflet, considered negligible, has not been taken into account. The described fluid dynamics, based on 2D or 3D CFD studies on local flow conditions, was modelled by means of a lumped parameterization, which leads to a second-order differential equation for each opening angle, $\vartheta$. Even though the adopted model for the valve motion is lumped, the equation for the dynamics of the opening angle, $\vartheta$, accounts for different physical mechanisms. Thus, global variations are modeled and in great part captured through the temporal variations of the valve area, $A$, and the opening angle, $\vartheta$. Fine details of the local dynamics–which are mostly influenced by the shape of the valve area–are not caught, thereby falling outside the goal of the present work. The angle $\vartheta$ reaches values in the range [$\vartheta_{min}$, $\vartheta_{max}$], where in healthy conditions $\vartheta_{min} = \vartheta_{min,h} = 0°$ (closed valve) and $\vartheta_{max} = \vartheta_{max,h} = 75°$ (fully open valve).

We related the valve area, $A$ [cm$^2$], to the opening angle, $\vartheta$, by means of the following law (Korakianitis & Shi, 2006):

$$A = \frac{(1-\cos\vartheta)^2}{(1-\cos\vartheta_{max,h})^2} A_h, \qquad (3)$$

where $A_h$ is the reference valve area value for an healthy adult. Only left-sided valvulopathies were investigated here, thus we set $A_h = 5$ cm$^2$ for the mitral valve and $A_h = 4$ cm$^2$ for the aortic valve (Baumgartner et al., 2009; Lancellotti et al., 2010a; Lancellotti et al., 2010b). In normal conditions, $A$ varies between 0 and $A_h$, with a quadratic dependence on $\vartheta$, as reported in Fig. 1 for the mitral (panel A) and aortic (panel B) valves.

## Grading left-sided valve disease severity

For each of the four left valvulopathies (AS, aortic stenosis, MS, mitral stenosis, AR, aortic regurgitation, MR, mitral regurgitation), we considered three valve area values, corresponding to different grades of severity (Baumgartner et al., 2009; Lancellotti et al., 2010a; Lancellotti et al., 2010b):

- AS: $A_s$ [cm$^2$] = 2 (mild), 1.25 (moderate), 0.90 (severe);
- MS: $A_s$ [cm$^2$] = 2 (mild), 1.25 (moderate), 0.90 (severe);
- AR: $A_r$ [cm$^2$] = 0.07 (mild), 0.20 (moderate), 0.33 (severe);
- MR: $A_r$ [cm$^2$] = 0.13 (mild), 0.30 (moderate), 0.44 (severe).

Observing the dependence between $A$ and $\vartheta$ introduced through Eq. (3), we expect lower $\vartheta_{max}$ values for increasing stenosis severity, and higher $\vartheta_{min}$ values for growing regurgitation grades.

For stenosis conditions, to find the maximum opening angle ($\vartheta_{max,s}$) corresponding to the stenotic area, $A_s$, we exploited Eq. (3) for each grade of severity as follows:

$$A_s = \frac{(1-\cos\vartheta_{max,s})^2}{(1-\cos\vartheta_{max,h})^2} A_h. \qquad (4)$$

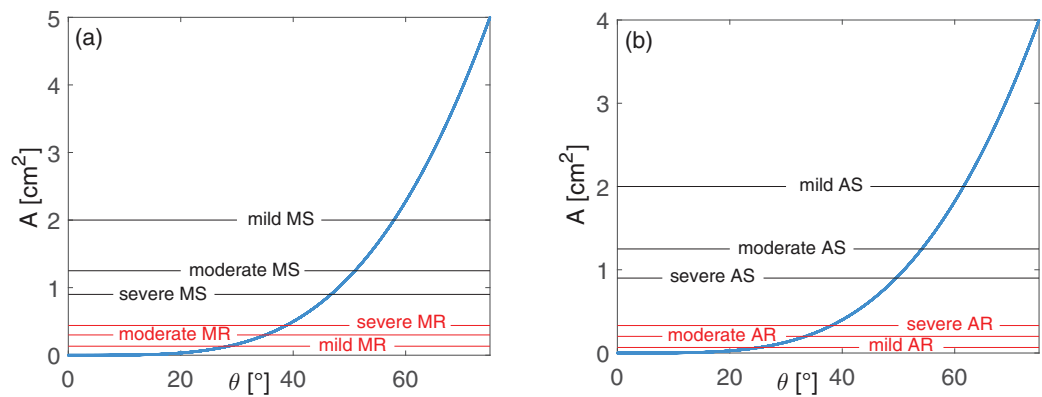

**Figure 1 Valve area *A* as function of the opening angle $\vartheta$: (A) mitral and (B) aortic valves.** Blue curves represent the healthy behavior, $A(\vartheta)$, as expressed by Eq. (3). Black horizontal lines represent $A_s$ values, while their intercepts with the blue curve individuate $\vartheta_{max,s}$ for different grades of stenosis, as formulated through Eq. (4). Red horizontal lines reproduce $A_r$ values, while their intercepts with the blue curve individuate $\vartheta_{min,r}$ for different grades of regurgitation, as expressed through Eq. (5).

In regurgitant conditions, the minimum opening angle ($\vartheta_{min,r}$) corresponding to the regurgitant orifice area, $A_r$, was found reformulating Eq. (3) as reported below:

$$A_r = \frac{(1 - \cos \vartheta_{min,r})^2}{(1 - \cos \vartheta_{max,h})^2} A_h. \quad (5)$$

From Eqs. (4) and (5) we were able to easily extract the opening angles $\vartheta_{max,s}$ and $\vartheta_{min,r}$ related to each grade of stenosis and regurgitation, respectively. A scheme summarizing the $\vartheta_{min}$ and $\vartheta_{max}$ values used in the model for the healthy and the twelve valve diseased configurations is provided in Table S1. Both stenosis and regurgitation were modelled in a simplified manner through geometrical variations of the opening angles $\vartheta$, accounting for the mechanical dysfunctions of the valve opening/closure failure. Because of the lack of clear data, during stenosis the increased stiffness of the leaflets is neglected, thus these latter were assumed as in healthy conditions. Altered valvular functions–due to valve prolapse, rheumatic disorders, congenital heart defects or endocarditis, and usually associated with regurgitation–were also not taken into account.

The proposed algorithm was used to simulate a specific grade of valvulopathy, once the corresponding reference valve area value is given. To double check the validity of this procedure, besides the hemodynamic parameters introduced at the beginning of this section, we also evaluated as post-processing parameters the regurgitant volumes, *RV* [ml/beat] (for regurgitations), and the mean pressure gradients, *MPG* [mmHg] (for stenosis), to evaluate the indexes recommended by current clinical guidelines to grade regurgitation and stenosis severity (*Baumgartner et al., 2009*; *Lancellotti et al., 2010a*; *Lancellotti et al., 2010b*). Recall that *RV* for both left valves was calculated as defined in Eq. (2). For *MPG* we used the velocity across the valve, $v = Q/A$ [m/s], and the Bernoulli equation, defining the transvalvular pressure gradient, $\Delta P = 4v^2$ [mmHg]. The mean pressure gradient, *MPG*, was calculated by averaging the instantaneous gradients, $\Delta P$, over the systolic phase (i.e., when there is forward flow $Q^+$) (*Baumgartner et al., 2009*). Mean pressure gradient, *MPG*, for stenosis and regurgitant volume, *RV* (as absolute values), for regurgitation, are reported in Table S2, as averaged over 5,000 cardiac periods.

### Simulations
To mimic AF conditions, both atria were assumed to be passive, i.e. atrial elastances were kept constant. A condition of lone AF was first simulated as reference baseline. Then, twelve simulations reproducing AF together with a specific grade of left valvulopathy were run. A ventricular contractile dysfunction has been described in both stenosis and regurgitation (*Maganti et al., 2010*), though without definitive results (*Shikano et al., 2003*). Given the lack of clear data (*Scarsoglio et al., 2014*) during heart valve diseases in AF, the reduced left ventricular inotropy was not modelled here and a normal left ventricular contractility was assumed for all the configurations. For each simulation, the transient dynamics were exceeded after 20 periods (*Scarsoglio et al., 2014*). Afterwards, 5,000 cardiac cycles were computed and recorded to account for a period lasting about one hour. This choice allowed the statistical stationarity of the results to be achieved. For all the cardiovascular variables and hemodynamic parameters, mean and standard deviation values were calculated.

AF beating features were approximated extracting uncorrelated RR from an Exponentially Gaussian Modified distribution (mean $\mu$ = 0.67 s, standard deviation $\sigma$ = 0.16 s, rate parameter $\gamma$ = 8.47 Hz), which is unimodal and describes the majority of AF cases (*Hennig et al., 2006*; *Scarsoglio et al., 2014*). The twelve AF with left-valvular disease simulations present the same AF beating features of the lone AF case. The defective valve opening/closure was added by varying $\vartheta_{max}$ and $\vartheta_{min}$ values according to the criteria discussed in the previous Section.

## RESULTS
Outcomes of the thirteen simulations (lone AF simulation, plus twelve AF with left-valvular disease simulations) are presented in terms of mean, $\mu$, and standard deviation, $\sigma$, values, as computed over 5,000 cardiac periods. The cardiovascular hemodynamic outcomes for stenosis and regurgitation are given in Tables 1 and 2, respectively. First columns of Tables 1 and 2 both display reference results of lone AF to facilitate the comparison. It is worth reading the above Tables also in terms of $c_v = \sigma/\mu$, which gives a normalized measure of the data dispersion. To better highlight the hemodynamic-based changes, results are first divided by valvulopathy, with focus on the most severe state. Representative time series of left atrial and ventricular volumes, together with the probability density functions of pulmonary vein pressure, $P_{pvn}$, and cardiac output (*CO*), are shown in Fig. 2 for severe aortic and mitral stenosis (black and red curves, respectively), and in Fig. 3 for severe aortic and mitral regurgitation (black and red curves, respectively). Lone AF results are reported in both figures as the baseline configuration (blue curves). A comparative framework of the diseases accounting for their grading is then proposed.

PeerJTable 1 Mean and standard deviation of computed variables during AF with concomitant left-sided valvular stenosis simulations. Lone AF computed values are also reported.

| | Lone AF | Aortic stenosis (AS) | | | Mitral stenosis (MS) | | |
|---|---|---|---|---|---|---|---|
| | | *Mild* | *Moderate* | *Severe* | *Mild* | *Moderate* | *Severe* |
| $P_{la}$ [mmHg] | 9.82 ± 0.82 | 9.70 ± 0.83 | 9.69 ± 0.83 | 9.73 ± 0.83 | 10.13 ± 0.65 | 11.07 ± 0.66 | 12.29 ± 0.71 |
| $P_{lv}$ [mmHg] | 47.64 ± 47.35 | 48.10 ± 48.58 | 49.71 ± 51.18 | 51.95 ± 54.67 | 46.69 ± 47.06 | 44.45 ± 44.89 | 41.29 ± 41.74 |
| $V_{la}$ [ml] | 62.80 ± 5.50 | 62.02 ± 5.56 | 61.93 ± 5.55 | 62.17 ± 5.53 | 64.86 ± 4.31 | 71.12 ± 4.39 | 79.24 ± 4.72 |
| $V_{lv}$ [ml] | 93.82 ± 28.39 | 93.15 ± 27.95 | 93.99 ± 27.45 | 95.55 ± 26.78 | 88.55 ± 26.69 | 82.41 ± 24.93 | 76.29 ± 23.20 |
| $V_{lves}$ [ml] | 58.71 ± 2.41 | 56.26 ± 1.74 | 56.12 ± 1.88 | 56.97 ± 2.09 | 58.11 ± 2.10 | 55.64 ± 1.81 | 52.21 ± 1.90 |
| $V_{lved}$ [ml] | 118.28 ± 6.19 | 116.49 ± 6.78 | 116.36 ± 6.69 | 116.99 ± 6.34 | 117.44 ± 8.86 | 111.63 ± 11.92 | 104.12 ± 13.07 |
| $P_{sas}$ [mmHg] | 100.39 ± 13.24 | 101.22 ± 13.13 | 101.13 ± 12.85 | 100.58 ± 12.50 | 99.27 ± 12.97 | 94.61 ± 12.09 | 87.91 ± 11.39 |
| $P_{sas,dias}$ [mmHg] | 82.56 ± 7.35 | 83.97 ± 7.94 | 84.44 ± 7.92 | 84.34 ± 7.67 | 81.40 ± 6.80 | 77.43 ± 5.67 | 71.82 ± 5.16 |
| $P_{sas,syst}$ [mmHg] | 120.94 ± 3.35 | 121.13 ± 3.52 | 121.18 ± 3.37 | 120.55 ± 3.22 | 119.61 ± 2.58 | 113.66 ± 2.86 | 105.56 ± 3.76 |
| $P_{pas}$ [mmHg] | 17.35 ± 4.30 | 17.30 ± 4.34 | 17.28 ± 4.33 | 17.27 ± 4.32 | 17.57 ± 4.25 | 18.15 ± 4.03 | 18.85 ± 3.79 |
| $P_{pvn}$ [mmHg] | 10.36 ± 0.61 | 10.25 ± 0.62 | 10.23 ± 0.62 | 10.26 ± 0.62 | 10.66 ± 0.58 | 11.57 ± 0.63 | 12.76 ± 0.68 |
| SV [ml] | 59.57 ± 7.74 | 60.23 ± 7.86 | 60.24 ± 7.90 | 60.02 ± 7.54 | 59.34 ± 9.65 | 55.99 ± 11.62 | 51.91 ± 12.36 |
| EF [%] | 50.15 ± 4.35 | 51.47 ± 4.13 | 51.54 ± 4.17 | 51.10 ± 4.00 | 50.17 ± 4.96 | 49.59 ± 5.64 | 49.14 ± 6.01 |
| CO [l/min] | 5.60 ± 1.16 | 5.66 ± 1.24 | 5.64 ± 1.15 | 5.61 ± 1.15 | 5.51 ± 1.20 | 5.24 ± 1.34 | 4.83 ± 1.26 |

Note:
CO, cardiac output; EF, ejection fraction; $P_{la}$, left atrium pressure; $P_{lv}$, left ventricular pressure; $P_{pas}$, pulmonary arterial pressure; $P_{pvn}$, pulmonary vein pressure; $P_{sas}$, systemic arterial pressure; $P_{sas,dias}$, diastolic systemic arterial pressure; $P_{sas,syst}$, systolic systemic arterial pressure; SV, stroke volume; $V_{la}$, left atrium volume $V_{lv}$, left ventricular volume; $V_{lved}$, left ventricular end-diastolic volume; $V_{lved}$, left ventricular end-systolic volume.

Table 2 Mean and standard deviation of computed variables during AF with concomitant left-sided valvular regurgitation simulations. Lone AF computed values are also reported.

| | Lone AF | Aortic regurgitation (AR) | | | Mitral regurgitation (MR) | | |
|---|---|---|---|---|---|---|---|
| | | *Mild* | *Moderate* | *Severe* | *Mild* | *Moderate* | *Severe* |
| $P_{la}$ [mmHg] | 9.82 ± 0.82 | 10.71 ± 0.90 | 11.99 ± 0.95 | 12.83 ± 0.93 | 11.08 ± 1.26 | 12.37 ± 1.76 | 13.20 ± 2.09 |
| $P_{lv}$ [mmHg] | 47.64 ± 47.35 | 48.05 ± 46.41 | 49.03 ± 45.32 | 49.79 ± 44.79 | 45.15 ± 43.75 | 41.77 ± 39.63 | 38.84 ± 36.52 |
| $V_{la}$ [ml] | 62.80 ± 5.50 | 68.73 ± 5.99 | 77.24 ± 6.31 | 82.86 ± 6.20 | 71.21 ± 8.43 | 79.83 ± 11.71 | 85.34 ± 13.93 |
| $V_{lv}$ [ml] | 93.82 ± 28.39 | 101.15 ± 34.79 | 112.25 ± 44.18 | 120.51 ± 50.65 | 97.23 ± 36.02 | 99.67 ± 44.03 | 100.74 ± 49.28 |
| $V_{lves}$ [ml] | 58.71 ± 2.41 | 57.90 ± 2.70 | 57.33 ± 2.46 | 57.22 ± 2.22 | 51.45 ± 2.41 | 42.36 ± 2.43 | 36.97 ± 1.75 |
| $V_{lved}$ [ml] | 118.28 ± 6.19 | 133.62 ± 8.04 | 159.13 ± 11.94 | 177.95 ± 13.26 | 130.22 ± 7.69 | 141.83 ± 9.25 | 148.96 ± 10.09 |
| $P_{sas}$ [mmHg] | 100.39 ± 13.24 | 93.31 ± 18.04 | 83.13 ± 25.20 | 76.15 ± 30.40 | 91.66 ± 13.07 | 82.96 ± 12.63 | 77.54 ± 12.00 |
| $P_{sas,dias}$ [mmHg] | 82.56 ± 7.35 | 69.23 ± 9.95 | 48.79 ± 12.03 | 35.09 ± 11.90 | 74.96 ± 7.38 | 67.57 ± 7.14 | 63.16 ± 6.73 |
| $P_{sas,syst}$ [mmHg] | 120.94 ± 3.35 | 119.36 ± 4.19 | 117.99 ± 3.50 | 117.79 ± 2.75 | 112.67 ± 3.22 | 104.33 ± 3.14 | 98.71 ± 3.14 |
| $P_{pas}$ [mmHg] | 17.35 ± 4.30 | 17.69 ± 4.06 | 18.18 ± 3.66 | 18.48 ± 3.41 | 17.94 ± 3.93 | 18.55 ± 3.56 | 18.96 ± 3.32 |
| $P_{pvn}$ [mmHg] | 10.36 ± 0.61 | 11.21 ± 0.64 | 12.43 ± 0.64 | 13.23 ± 0.60 | 11.57 ± 0.88 | 12.82 ± 1.17 | 13.61 ± 1.38 |
| SV [ml] | 59.57 ± 7.74 | 75.72 ± 10.04 | 101.80 ± 13.73 | 120.73 ± 14.66 | 78.76 ± 8.98 | 99.48 ± 10.27 | 112.00 ± 10.59 |
| EF [%] | 50.15 ± 4.35 | 56.41 ± 4.44 | 63.68 ± 4.12 | 67.59 ± 3.56 | 60.28 ± 3.79 | 69.95 ± 3.24 | 75.03 ± 2.47 |
| CO [l/min] | 5.60 ± 1.16 | 5.27 ± 1.50 | 4.80 ± 2.18 | 4.45 ± 2.46 | 5.13 ± 1.26 | 4.65 ± 1.34 | 4.34 ± 1.34 |

Note:
For the abbreviations, please refer to Table 1.

### Stenosis

During AS, data dispersion remained practically unvaried with respect to lone AF, with the only exception of $P_{lv}$, presenting more dispersion. An increased mean $P_{lv}$ value is a consequence of the higher aortic resistance during AS and is necessary to guarantee an adequate $CO$. Moreover, volume time series (Figs. 2A and 2B) and probability density functions (Figs. 2C and 2D) preserved the same behavior and shape as observed during lone AF, thereby confirming the modest hemodynamic impact of AS already evidenced by data dispersion.

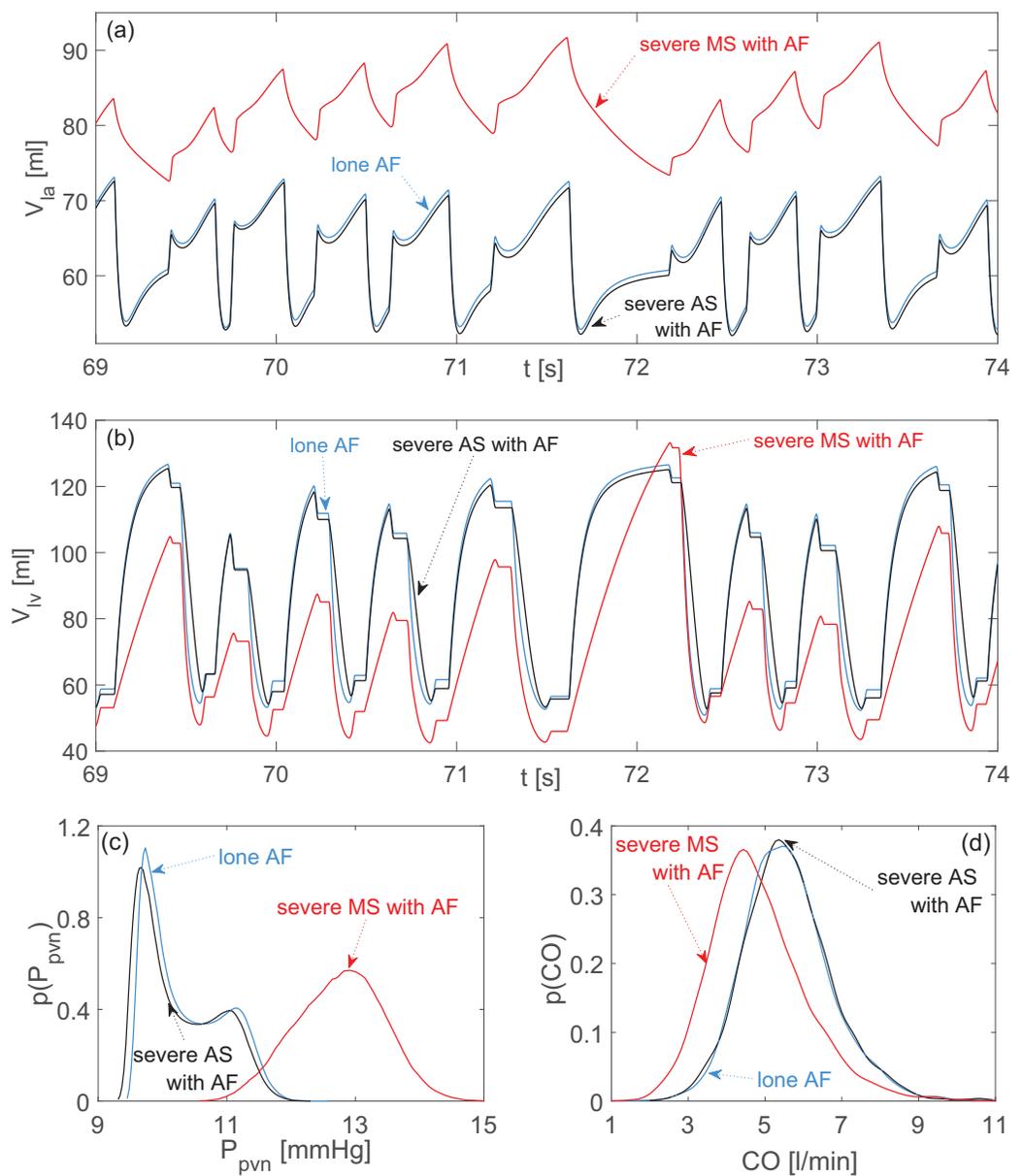

**Figure 2 Aortic and mitral stenosis with AF compared to lone AF.** Representative time series (the same stochastic $RR$ series is used for the three configurations): (A) left atrial volume, $V_{la}$; (B) left ventricular volume, $V_{lv}$. Probability density functions: (C) pulmonary vein pressure, $P_{pvn}$; (D) cardiac output, $CO$. Blue curves: lone AF. Black curves: severe aortic stenosis with AF. Red curves: severe mitral stenosis with AF.

The scenario was different for MS. With respect to lone AF, dispersion of data decreased for atrial variables ($P_{la}$ and $V_{la}$), $P_{pvn}$ e $P_{pas}$, while performance indexes experienced more

PeerJ

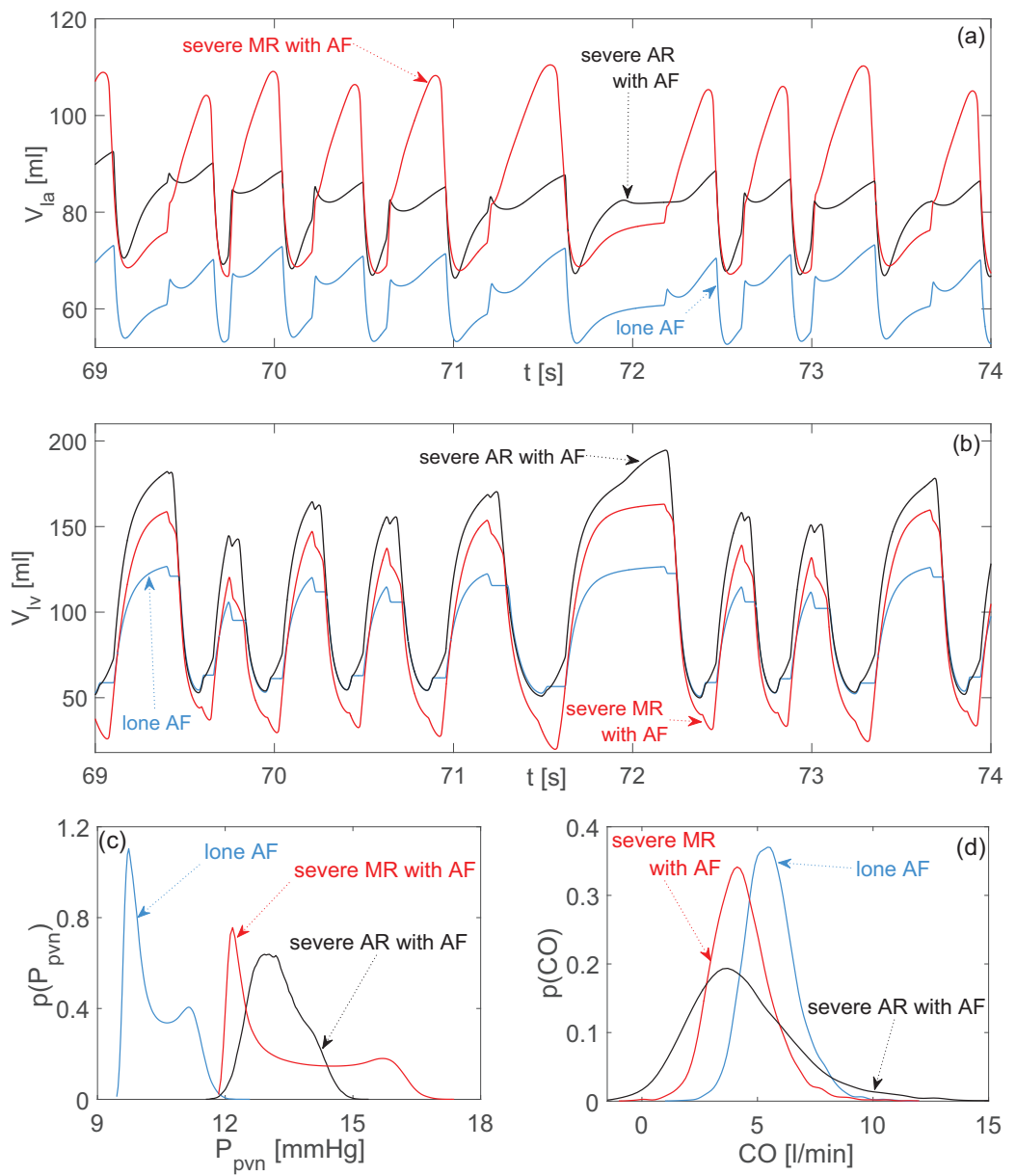

**Figure 3 Aortic and mitral regurgitation with AF compared to lone AF.** Representative time series (the same stochastic *RR* series is used for the three configurations): (A) left atrial volume, $V_{la}$; (B) left ventricular volume, $V_{lv}$. Probability density functions: (C) pulmonary vein pressure, $P_{pvn}$; (D) cardiac output, *CO*. Blue curves: lone AF. Black curves: severe aortic regurgitation with AF. Red curves: severe mitral regurgitation with AF.

dispersion (*SV, CO, EF*). Atrial overload is detectable by the increased mean $V_{la}$ and $P_{pvn}$ values, as well as by the different shape assumed by the $V_{la}$ time series and the $P_{pvn}$ probability density function with respect to lone AF (Figs. 2A and 2C). Changes at ventricular level were less pronounced, but largely imputable to inefficient atrial ejection. This latter in turn reduced $V_{lved}$ values, leading to an overall *SV* reduction. The cardiac efficiency, *CO*, was weakened as a result of the decreased mean net volume available to be ejected from ventricle to the aorta.

## Regurgitation

Both aortic and mitral regurgitation similarly increased the mean atrial volume. However, MR induced the highest peak values (up to 110 ml) and substantially changed the temporal dynamics with respect to lone AF (Fig. 3A). The enlarged atrial volume led for both regurgitations to an increase of $P_{pvn}$, with an accentuated right tail for the probability density function of MR (Fig. 3C).

In case of AR, data dispersion decreased for atrial variables, $P_{pvn}$, $P_{pas}$, $P_{lv}$, EF, with respect to lone AF, while data were sparser for $P_{pas}$, CO, $V_{lv}$. The failed closure of the aortic valve during diastole caused substantial regurgitant flow from the aorta back to the ventricle. This regurgitation on the one hand promoted ventricular overfilling, with elevated $V_{lved}$ values (Fig. 3B), which in turn partially inhibited the normal atrial emptying. On the other hand, the regurgitant flow reduced the net antegrade CO, into the aorta (Fig. 3D).

Comparing MR with respect to lone AF, data dispersion was lower for $P_{lv}$, $P_{pas}$, SV and EF, while it increased for atrial variables, $P_{pvn}$, $V_{lv}$, and CO. The defective closure of the mitral valve during systole resulted in regurgitant flow from ventricle towards the atrium, causing high $V_{la}$ peaks and abnormally emptying of the ventricle after ejection (i.e., decrease of $V_{lves}$, Fig. 3B). As a consequence, the net forward CO, was reduced (Fig. 3D). At the end of systole, the atrium was overfilled and ejected a greater amount of blood into the ventricle during diastole, leading eventually to an increase of $V_{lved}$.

## Comparative framework of valvular heart disease

Recall that dispersion of data is mainly produced by irregular beating. Changes in the dispersion of the results–with respect to lone AF–can be interpreted as the (more or less) pronounced ability of the valvulopathy to modify AF hemodynamics. From this point of view, AS had the least impact since dispersion remains basically unaltered, while both MR and AR acted to substantially vary the cardiovascular response.

In order to compare the relative effects of each valvular disease by grade, the percentage variation of every averaged hemodynamic variable compared to the control, lone AF simulation, was evaluated. Figure 4 shows the most significant percentage variations, involving atrial and upstream pulmonary venous return (A), ventricular dynamics (B and C), performance indexes (D and F), and systemic arterial pressure (E). In the pulmonary circulation, although mean pulmonary arterial pressure ($P_{pas}$) did not undergo substantial changes, mean pulmonary vein pressure ($P_{pvn}$) increased by 31.4, 27.7, and 23.2%, in case of severe MR, AR, and MS, respectively (Fig. 4A). Similarly, mean left atrial pressure ($P_{la}$), increased by 34.4, 30.7 and 25.2% in the cases of severe MR, AR and MS, respectively. In the left ventricle, an increase in mean left ventricular pressure ($P_{lv}$) was seen in severe AS (+9.0%), while there was a decrease in severe MS (−13.3%) and MR (−18.5%) (Fig. 4B); mean left ventricular volume ($V_{lv}$) increased due to severe AR (+28.8%) and MR (+7.4%), and decreased in case of severe MS (−18.7%) (Fig. 4C). Concomitantly, stroke volume (SV) showed an upsurge in severe AR (+102.7%) and MR (+88.0%), and a decrease due to severe MS (−12.9%) (Fig. 4D). Finally, mean systemic arterial pressure ($P_{sas}$) declined in severe AR (−24.1%), MR (−22.8%) and MS (−12.4%) (Fig. 4E), with an analogous decrease in CO in severe MR (−22.5%), AR (−20.5%) and MS (−13.8%) simulations (Fig. 4F).

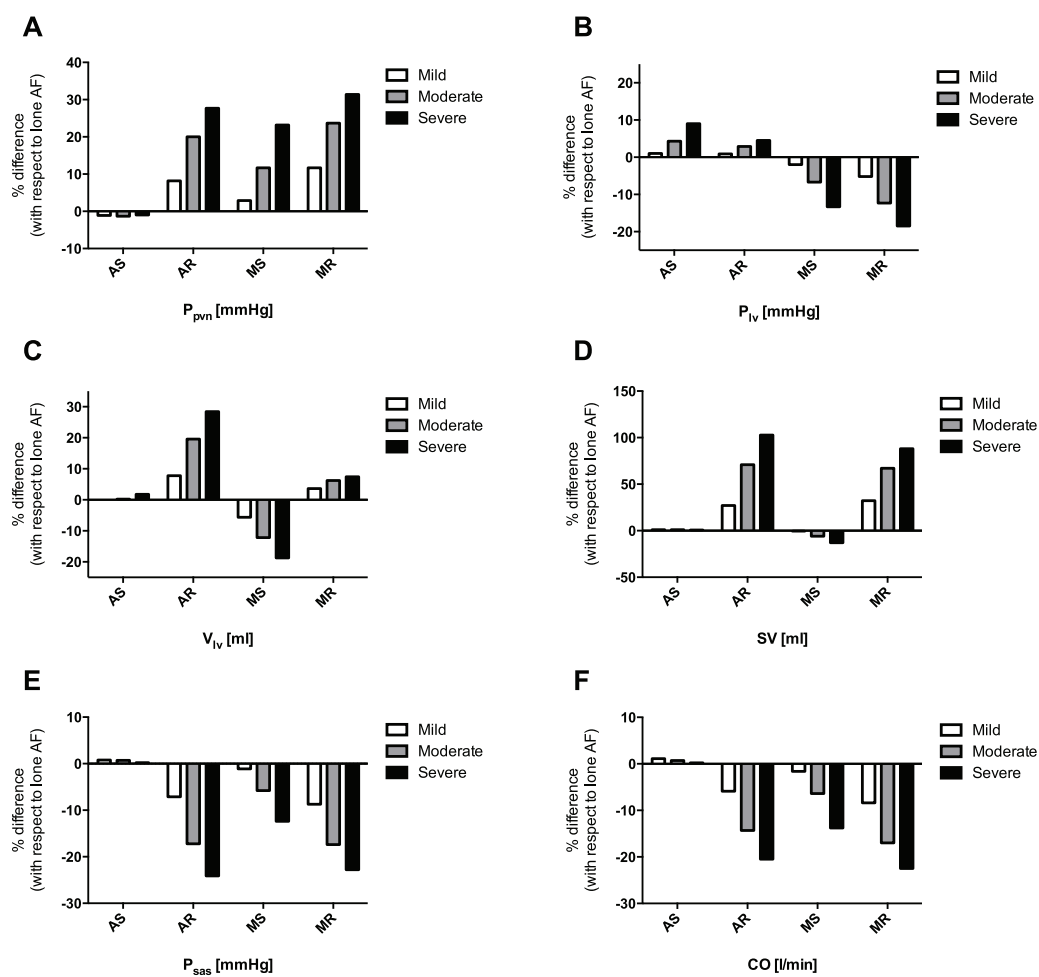

**Figure 4 Grouped plot displaying percentage variations, referred to lone AF simulation, of selected computed variables for each concomitant valvular disease.** (A) $P_{pvn}$, (B) $P_{lv}$, (C) $V_{lv}$, (D) $SV$, (E) $P_{sas}$, (F) $CO$.

## DISCUSSION

The present study focused on computationally assessing the hemodynamic impacts exerted by different left-sided valve diseases in the context of persistent AF. Previous literature has not addressed this particular topic, which warrants attention given the substantial proportion of AF patients presenting with concomitant valvular heart disease. Indeed, AF frequently complicates mitral valve diseases (MS and MR), especially when their etiology is rheumatic. In aortic valve diseases, AF has been less well studied, but it often complicates uncorrected AS or AR (*Darby & DiMarco, 2012*; *Vora, 2006*).

To simulate AF in the context of different left-sided valve diseases, we used a lumped model of the cardiovascular system previously validated for lone AF (*Scarsoglio et al., 2014*). This model has two fundamental features: (i) the ability to simulate persistent AF;

(ii) a detailed description of valve dynamics, allowing the modelling of different valvulopathies. In fact, as detailed in the Materials & Methods Section, by developing an innovative algorithm to model precise severity grades for each valve disease, we were able to predict hemodynamic variables for each valvular disease, grading the proportional variation compared to the lone AF simulation. In general, the valvulopathy disease grading design proved appropriate and reproducible when compared to clinically used indexes: the calculations of mean pressure gradients across the valve for stenosis and regurgitant volumes for regurgitation (Table S2) yielded results in agreement with the ranges indicated by current guidelines (Baumgartner et al., 2009; Lancellotti et al., 2010a; Lancellotti et al., 2010b). A proper modelling of the ventricular inotropy (here neglected) is expected to reduce, especially for severe grades of valvular diseases, the systemic and ventricular pressures as well as the severity indexes (*MPG* for stenosis and *RV* for regurgitation), which are now, therefore, plausibly overestimated. In this setting, though lacking the presence of autonomic nervous system effects, the model allows one to simulate the cardiovascular system at a "steady-state" without autonomic influence, thus highlighting the pure hemodynamic component that each valve disease exhibits during AF.

During AF, based on the current computational analysis, MR and AR had the strongest impact on hemodynamics, followed by MS; conversely, AS had by far the least impact among the studied valvular diseases. In particular, MR displayed the most influence at the level of the left atrium and in the upstream pulmonary circulation, as indicated by increased $P_{la}$ and $P_{pvn}$ (Fig. 4A), together with a strong impairment in $P_{sas}$ and *CO* (Figs. 4E and 4F), due to the regurgitating blood volume into the atrium. AR resembled MR hemodynamically but with more impairment in *CO*. The MS effects during AF, although relevant, were less pronounced than either regurgitation, either on left atrium/pulmonary circulation or on $P_{sas}$ and *CO*. Finally, in the case of AS, only a small rise in $P_{lv}$ (Fig. 4B) was seen. For all the other hemodynamic parameters, AS did not show any detectable trend when shifting from mild to severe grades, while the other valvulopathies clearly did.

From a fluid dynamics point of view, we can try to untangle why regurgitation was hemodynamically more problematic than stenosis, considering that the latter makes peak forward flow rate slow and inefficient because of a higher outflow resistance, though no substantial flow directional variation is introduced with respect to the nonstenotic state. Changes in flow direction can be quantified by means of the regurgitant volume, *RV*. For all grades of both aortic and mitral stenosis, *RV* absolute mean values did not exceed 6 ml/beat, falling within the physiologic range (Scarsoglio et al., 2016). Regurgitation led instead to a drastic change in flow direction (please refer to the *RV* values in Table S2) which, in the presence of normal valve closure, had no counterpart in healthy dynamics. As vortex effects play an important role in valve motion (Korakianitis & Shi, 2006), it can reasonably be expected that their dynamics can be affected when a significant portion of fluid regurgitates backward.

Moreover, our data demonstrated that mitral valvulopathies are in general more hemodynamically disruptive than aortic ones for the following reasons. In the case of aortic valve disease, proper functioning of the mitral valve was able to smooth and damp out the upstream changes (at the atrial level and proximally). When instead a mitral valvulopathy occurred, it directly involved the atrium, a region which already suffered from contractile dysfunction induced by AF. The mitral valve lost its regulating capability, thus hemodynamic variations almost equally affected atrial and ventricular regions, also influencing the upstream pulmonary venous return (e.g., $P_{pvn}$) and the downstream systemic arterial variables (e.g., $P_{sas}$).

The impact of increasing severity of valvulopathy varied considerably with the lesion. Mild MS resulted in very little hemodynamic disturbance, only becoming significant with higher grades of stenosis. In contrast, even milder forms of AR and MR were significant in the presence of AF. As an example, compared to the control values of lone AF, $P_{pvn}$ increased by 11.7% in mild MR and by 31.4% in severe MR (i.e. a nearly three-fold increase from mild to severe MR), while it underwent an increase of 2.9% in mild MS and 23.2% in severe MS (i.e., an eight-fold increase from mild to severe MS), suggesting that, although there is adaptation at lower grades, at the severe stage, MS has an impact of similar magnitude to regurgitation. A likely explanation for this behavior is the absence of atrial contraction in AF. Often referred to as the "atrial kick," atrial contraction, when present, can partially dampen the effects of MS when the grade of the disease is low.

**Limitations**

In addition to the previously stated lack of autonomic nervous system regulation, some other limitations of the present modelling study should be considered. First, AF conditions were set the same for all simulations in the attempt to quantify the "net impact" of the specific valve disease during the arrhythmia, regardless of other differential compensatory mechanisms that may, in fact, be present in clinical practice. Second, coronary circulation was not taken into account, since its peculiar features (e.g., diastolic flow) makes the modelling challenging; therefore, the effect of AF and different valve diseases on pressures and volumes in that circulation was not accounted for by the present model. Third, the model predicted hemodynamic effects of valvular disease during AF, without considering other pathological conditions, such as hypertension or heart failure, that could themselves affect cardiovascular variables. Moreover, linear relations are assumed for the pressure-volume constitutive equations in the vasculature, which can lead to an underestimation of diastolic pressures in severe stenosis conditions. In the end, AF beating features were limited to the unimodal distribution only, while multimodal RR distributions were not analyzed.

## CONCLUSIONS

The present study, based on a validated computational cardiovascular model for lone AF, provides new insights into the consequences of left-sided valvular disease with concomitant persistent AF, and elucidates which valvular diseases exert the worst hemodynamic effects. In general, valvular regurgitation had the strongest impact on hemodynamics, immediately followed by MS. Conversely, AS had the least impact among the studied valvular diseases. The present findings warrant further clinical investigation because, if confirmed, they may potentially impact AF management (for example, requiring the adoption of more aggressive rhythm control strategies, such as AF transcatheter ablation) in case of a specific valvular pathology.

## ACKNOWLEDGEMENTS
The authors would like to thank Mark Miller for his valuable contributions to the editing of the manuscript, and the reviewers, Gianni Pedrizzetti and Thomas Christian Gasser, for their constructive comments and suggestions which helped to improve the work.

## ADDITIONAL INFORMATION AND DECLARATIONS

### Funding
The authors received no funding for this work.

### Competing Interests
The authors declare that they have no competing interests.

### Author Contributions
- Stefania Scarsoglio conceived and designed the experiments, performed the experiments, analyzed the data, contributed reagents/materials/analysis tools, wrote the paper, prepared figures and/or tables, reviewed drafts of the paper.
- Andrea Saglietto conceived and designed the experiments, analyzed the data, wrote the paper, prepared figures and/or tables, reviewed drafts of the paper.
- Fiorenzo Gaita conceived and designed the experiments, analyzed the data, wrote the paper, reviewed drafts of the paper.
- Luca Ridolfi conceived and designed the experiments, analyzed the data, contributed reagents/materials/analysis tools, wrote the paper, reviewed drafts of the paper.
- Matteo Anselmino conceived and designed the experiments, analyzed the data, wrote the paper, reviewed drafts of the paper.

### Data Deposition
The following information was supplied regarding data availability:
Data sets and code scripts are available at Figshare.
DOI: 10.6084/m9.figshare.3465407;
https://figshare.com/articles/PeerJ2016_Scarsoglio/3465407.

### Supplemental Information
Supplemental information for this article can be found online at http://dx.doi.org/10.7717/peerj.2240#supplemental-information.